\documentclass[12pt,a4paper]{JHEP3}

\usepackage{graphicx}
\usepackage{amssymb,amsmath}
\usepackage{dsfont}
\usepackage{slashed}
\usepackage{braket}
\usepackage{epsfig}
\usepackage{subfigure}
\usepackage[percent]{overpic}

\title{Series of Tests of a Constraint on Asymptotically Free Gauge Theories}

\preprint{ }
\author{
 Yair Mulian\\\\
 Physics Department, Ben-Gurion University of the Negev,
 Beer Sheva 84105, Israel\\ \\
 }
\abstract{

In [1] Thomas Appelquist, Andrew G. Cohen, and Martin Schmaltz (ACS) proposed a constraint on the structure of asymptotically-free field theories. This constraint limits the number of degrees of freedom of asymptotically-free gauge theories in the infrared ($IR$) region relative to those in the ultra-violet ($UV$) region. 

In their paper ACS checked various examples, both supersymmetric and non-supersymmetric, but checked only one case of interacting $IR$ fixed point with superpotential - the case of Seiberg-dual of $SU(N_{c})$ gauge theory.

Here we will verify the conjecture for two new cases - $SO(N_{c})$ and $Sp(2N_{c})$ gauge groups around the Banks-Zaks fixed points.

In addition, we subject the ACS inequality to a series of nontrivial tests in theories with conjectured accidental symmetries in the $IR$ and dramatically different dynamics caused by superpotential deformations. We start with $ADE$-type deformations then move on to check three chiral theories by estimating their decoupled invariants from the chiral ring using a-maximization and unitarity considerations. 

Remarkably, we found no violation of the ACS conjecture.
}

\begin{document}

\bibliographystyle{JHEP}
\section{Introduction and Summary}

The prevailing point of view is that the theories describing nature's fundamental forces are a consequence of gauge principles and known as gauge theories. Perturbative techniques can be applied when we work with weak coupling (as for instance, in QED) but as we move to the strong coupling case (such as the low energy region of QCD) we cannot apply them anymore. One can still use the techniques of instantons, lattice approximation, etc. but instantons are not always controllable, while the lattice approach is numerical in nature and still cannot address many of the hard questions.  

One of the principles that may lead us to interesting insights on strongly coupled theories is the principle of {\it Seiberg duality} [4] [5]. This principle relates the correlation functions of a supersymmetric strongly coupled theory to a weakly coupled theory with a different gauge group and additional fields.
Using duality we avoid solving the dynamics at the strong coupling region and instead rely on symmetries, inspired guesswork, and general properties of supersymmetry.

In 1999 a new constraint on the structure of strongly coupled gauge theories was proposed by ACS [1]. This was not the first time that inequalities on the structure of gauge theories appeared. For instance, the c-theorem (known also as Zamolodchikov's theorem) was proposed in 1986 concerning the RG flow of field theories in the case of two dimensions [3].
Finding the ``four-dimensional analog" to the c-theorem was not an immediate task, and the a-theorem was proposed by Cardy in 1988 [2].
The a-theorem was proved lastly in [14] [15] but until now no proof is known for the ACS proposal, neither there is a couterexample. The state of the art is that it is known to hold for many nontrivial examples (including perturbative fixed points, SUSY gauge theories, etc.). 

We will now define the ACS conjecture more precisely. There is no single way to define the number of degrees of freedom, but ACS in their paper choosed to work with quantities which are related to the free energy of the field theory (only theories in which the free energy may be rendered finite and cutoff independent will be used here) and measure them in the $IR$ and $UV$ region by $$f_{IR}\equiv -lim_{T\rightarrow0}\frac{90\mathcal{F}}{\pi^{2}T^{4}}$$ and $$f_{UV}\equiv -lim_{T\rightarrow\infty}\frac{90\mathcal{F}}{\pi^{2}T^{4}} $$ respectively (where $\mathcal{F}$ denotes the free energy and $T$ the temperature).
Utilizing these definitions, ACS conjectured the inequality $f_{IR}\leq f_{UV} $ \footnote{It should be mentioned that for a general field theory in more than two dimensions the free energy as a function of $f(T)$ does not interpolate monotoincally between $f_{IR}$ and $f_{UV}$ [1].}.

There are many motivations for studying this inequality along with the other constraints on RG flows. It turns out that the ACS constraint can shed light on the phase diagram of a theory. For example, in [1] it was shown that the free energy leads to the conformal window in a manner consistent with that known from Seiberg duality. The hope is that by proving the ACS conjecture we can apply this result for QCD and establish a rigorous bound on the conformal window.

The structure of this paper will split into two parts: In the first section we will verify ACS on the Banks-Zaks fixed point for SQCD with $SO(2N_{c})$ and $Sp(2N_{c})$ gauge groups in a similar way to [1].

Another type of examinations will be continued in the section afterwards. We begin by explaining the unitarity bound and the procedure described by David Kutasov, Andrei Parnachev and David A. Sahakyan [7] on how to correct the central charge in case of unitarity violation (a-maximization). Using this procedure we can estimate the degrees of freedom in the $IR$ as we consider each violation to be originated from particles becoming free. We will start by focusing mainly on non-chiral $SU(N_{c})$ gauge group theories and describe the possible ways to deform it by introduction of superpotential composed of one or two adjoint fields (known as $ADE$ classification [9]), then move on to theories with semi-simple gauge group and chiral theories with and without a superpotential.

In summary, our objective was to check possible violations of the ACS conjecture. The result we found is that the ACS conjecture holds for each of the theories checked, providing strong basis for its general validity.

\section{Checking ACS in SQCD with an Interacting Infrared Fixed Point}
Following $[1]$ the appendix contains three exmples in which the theory is both UV free and having a free magnetic dual. The ACS inequality cease to work in these examples exactly where the IR dual is no longer free and our underlying assumption fails. In order to check the inequality also in the interacting region we use the perturbative expansion for the free energy of high-temperature gauge theory [11][12][13]:
\begin{equation}f(T)=f_{UV}-\frac{45}{32\pi^{2}}N_{g}(C(G)+3T(\Phi))g^{2}+...\end{equation}
where $C(G)\delta^{AB}=f^{ACD}f^{BCD}$  ($f^{ABC}$ denotes the structure constant of the group $G$) is the quadratic Casimir of the adjoint representation and $T(\Phi)=\sum_{i=1}^{i=N_{f}}T_{i}(\Phi)$ , $T_{i}(\Phi)\delta^{AB}=TrR^{A}R^{B}$ ($R^{A}$ are the generators of the gauge group) is the trace normalization factor, lastly, $N_{g}$ denotes the number of gauge bosons.

The free energy in the interacting IR region can be found by using the Seiberg duality and adding the contribution that comes from the Yukawa sector. While the UV theory is still UV-free the IR dual now flows to non-zero values of the couplings (named fixed points). The formula to get the fixed point coupling values is found in [10]:
\begin{equation}\begin{split}&\frac{dg}{dt}=\frac{g^{3}}{16\pi^{2}}(T(\Phi)-3C(G))+\frac{g^{5}}{(16\pi^{2})^{2}}(-6(C(G))^{2}+2C(G)T(\Phi)+4T(\Phi)C(\Phi))-\\
&\frac{g^{3}}{(16\pi^{2})^{2}}Y^{ijk}Y^{ijk}\frac{C(\Phi_{k})}{d(G)}\end{split}\end{equation}
where the coupling $Y^{ijk}$ are the coefficients of the cubic part of the superpotential:
\begin{equation}W=\frac{1}{6}Y^{ijk}\Phi^{i}\Phi^{j}\Phi^{k}+\frac{1}{2}\mu^{ij}\Phi_{i}\Phi_{j}+L^{i}\Phi_{i}\end{equation}
and the value of the Yukawa coupling at the fixed point is found from [10]:
\begin{equation}\frac{dY^{ijk}}{dt}=Y^{ijp}\left[\frac{1}{16\pi^{2}}\gamma_{p}^{(1)k}+\frac{1}{(16\pi^{2})^{2}}\gamma_{p}^{(2)k}\right]+(k\leftrightarrow i)+(k\leftrightarrow j)\end{equation}
\newline where \newline
$\gamma_{i}^{(1)j}=\frac{1}{2}Y_{ipq}Y^{jpq}-2\delta_{i}^{j}g^{2}C(\Phi_{i})$
\newline
$\gamma_{i}^{(2)j}=-\frac{1}{2}Y_{imn}Y^{npq}Y_{pqr}Y^{mrj}+g^{2}Y_{ipq}Y^{jpq}[2C(\Phi_{p})-C(\Phi_{i})]+2\delta_{i}^{j}g^{4}[C(\Phi_{i})S(R)+2C(\Phi_{i})^{2}-3C(G)C(\Phi_{i})] $\\

In general expressions $(2.2)$ and $(2.4)$ are not enough to find the fixed point as there are contributions that comes with higher orders of $g$. In order for this expansion to be suffice we take $N$ very large, that will make all higer orders contributions decay to zero [8] [38] (this type of fixed points known as Banks-Zaks fixed points). Inserting all the parameters into the above foumulas we find the following fixed point for $SO(N_{c})$ group with $N_{f}$ flavors:
\begin{equation}g^{2}=\frac{14}{3}\frac{\epsilon}{N_{c}} \qquad\qquad y^{2}=\frac{2}{3}\frac{\epsilon}{N_{c}}\end{equation}
while for $Sp(2N_{c})$ group with $2N_{f}$ flavors we find:
\begin{equation}g^{2}=\frac{7}{9}\frac{\epsilon}{N_{c}} \qquad\qquad y^{2}=\frac{2}{3}\frac{\epsilon}{N_{c}}\end{equation}
where $\epsilon\equiv\frac{2N_{f}-3N_{c}}{N_{c}}$ and we see that in order to have the perturbative expansion valid we must take $0<\epsilon\ll1$. With this data we are now going to check the ACS inequality.\newline 
\subsection{$SO(N_{c})$ group with $N_{f}$ flavors}
In the case we are interested expression $(2.1)$ becomes:
\begin{equation}f(T)=f_{UV}-N_{c}(N_{c}-1)(N_{c}+3N_{f}-2)\frac{45g^{2}}{128\pi^{2}}+...\end{equation}
and the Seiberg dual is found by making the transformation $N_{c}\rightarrow N_{f}-N_{c}+4$ and adding $\frac{N_{f}(N_{f}+1)}{2}$ mesons:
\begin{equation}f(T)=f_{IR}-(N_{f}-N_{c}+3)(N_{f}-N_{c}+4)(4N_{f}-N_{c}+2)\frac{45g^{2}}{128\pi^{2}}-3N_{f}^{2}(N_{f}-N_{c}+4)\frac{45y^{2}}{64}\end{equation}
On the Banks-Zaks fixed point $(2.5)$ the last expressions becomes:
\begin{equation}f_{UV}=\frac{15}{2}N^{2}(1+\frac{\epsilon}{4})\end{equation}
\begin{equation}\ensuremath{f_{IR}=\frac{15}{2}N^{2}(1+\epsilon)-\frac{525}{16}N^{2}\epsilon-\frac{405}{16}N^{2}\epsilon=}\frac{15}{2}N^{2}(1+\epsilon-\frac{31}{2}\epsilon)\end{equation}
Clearly, we can see that $f_{UV}>f_{IR}$.
\newline\newline
\subsection{$Sp(2N_{c})$ group with $2N_{f}$ flavors}
The expressions for the free energy in the $UV$ is:
\begin{equation}f(T)=f_{UV}-N_{c}(2N_{c}+1)(N_{c}+3N_{f}+1)\frac{45g^{2}}{32\pi^{2}}+...\end{equation}
while the $IR$ free energy expression is found by applying the Seiberg prescription $2N_{c}\rightarrow2N_{f}-2N_{c}-4$ and adding $N_{f}(2N_{f}-1)$ mesons we find:
\begin{equation}\begin{split}&f(T)=f_{IR}-(N_{f}-N_{c}-2)(2N_{f}-2N_{c}-3)(4N_{f}-N_{c}-1)\frac{45g^{2}}{32\pi^{2}}\\
&-3N_{f}^{2}(N_{f}-N_{c}+2)\frac{45y^{2}}{32}\end{split}\end{equation}
Inserting the Banks-Zaks fixed point $(2.6)$ we find again that $f_{UV}>f_{IR}$.
\section{Checking ACS using the a-theorem}
The characters of the conformal superalgebra was investigated in many places (see for example [28]). We can classify the unitarity irreducible representations of an operator $\mathcal{O}$ in 4-dimensional space-time by using a real number $\Delta $, the exact scaling dimension, defined for spinless field by:
\begin{equation}
\phi(x)\rightarrow\phi(x^{\prime})=\lambda^{\Delta}\phi(0)
 \end{equation}
and two additional indices $(j_{L},j_{R})$ which denotes the irreducible representation of the Lorentz group $SL(2\mathbb{C})$.
Of special interest is a conformal primary operator, which transforms under conformal gage transformations as 
\begin{equation}
\phi(x)\rightarrow\left|\frac{\partial x^{\prime}}{\partial x}\right|^{\frac{\Delta}{d}}\phi^{\prime}(x^{\prime})
\end{equation}
where $\left|\frac{\partial x^{\prime}}{\partial x}\right|$ is the Jacobian of the transformation of the conformal transformation of the coordinates.

Unitarity implies bounds on the representations as we need to require that all the states in a representation have positive norm (states which break unitarity corresponds to representations with null states which can be removed leaving a shorter representation). Using the $(j_{R},j_{L}) $ Lorentz representation we can write for a primary operator $\mathcal{O}_{(j_{L},j_{R})}$:
$$\Delta\geq j_{L}+j_{R}+2, \qquad j_{L}\cdot j_{R}\neq0$$
$$\Delta\geq j_{L}+j_{R}+1, \qquad j_{L}\cdot j_{R}=0$$
A scalar operator, $\mathcal{O}$,  saturates the second bound occurs iff the operator satisfies the Klien-Gordon equation:
$$\partial^{2}\mathcal{O}=0\qquad\Longleftrightarrow\qquad\Delta(\mathcal{O})=1$$
Similarly a spin-$\frac{1}{2}$ operator, $\mathcal{O}^{\dot{\alpha}}$,  saturates the unitarity bound iff it satisfies Dirac equation:
$$\sigma_{\alpha\dot{\alpha}}^{\mu}\partial_{\mu}\mathcal{O}^{\dot{\alpha}}=0\qquad\Longleftrightarrow\qquad\Delta(\mathcal{O})=\frac{3}{2}$$
And an anti-symmetric tensor, $\mathcal{O}_{[\mu\nu]}$, satisfies the bound only iff it satisfies Maxwell equation:
$$\partial^{\mu}\mathcal{O}_{[\mu\nu]}=0\qquad\Longleftrightarrow\qquad\Delta(\mathcal{O}_{[\mu\nu]})=\frac{5}{2}$$
Whether or not the theory is conformal, we can express the unitarity bound also by using the $U(1)_{R}$ symmetry charge of the operator $\mathcal{O}$ [8]:
$$\ensuremath{\Delta(\mathcal{O})\geq\frac{3}{2}|R(\mathcal{O})|} $$
where saturation of the last bound occurs for primary chiral operators (as defined in [9]), then:
$$R(\mathcal{O})=\frac{2}{3}\Delta(\mathcal{O})=\frac{2}{3}(1+\frac{1}{2}\gamma)$$
where $\gamma_{i}$ is the anomalous dimension of the field.

The unitary bound, $\Delta(\mathcal{O})\geq1$ for spin zero operators, along with the last inequality implies that any operator should also satisfy $\ensuremath{R(\mathcal{O})}\geq\frac{2}{3} $, with $\ensuremath{R(\mathcal{O})}=\frac{2}{3} $ if and only if the operator $\mathcal{O}$ is a decoupled free field with $\partial_{\mu}\partial^{\mu}\mathcal{O}=0 $.\newline

As explained in [8], this unitarity condition actually does not constrain the R-charge assignment of the fields inside the Lagrangian, and indeed (as we are about to see) it can happen that the R-change of a field appears to violate the $\ensuremath{R(\mathcal{O})}\geq\frac{2}{3} $ bound.
The solution to this apparent conflict is that any gauge invariant chiral operator $X$ which apparently violate the unitarity condition is actually a free decoupled field and have an accidental extra $U(1)_{X}$ symmetry to correct the superconformal R-charge to be $R(X)=\frac{2}{3}$, with the R-charge of other operators unaffected.

The way to determine the ``real" R-charge is also explained in [8]. Their idea was to write down the most general R-symmetry possible:
$$R_{t}=R_{0}+\sum_{I}s_{I}F_{I} $$ 

Where $F_{I}$ are all the non-R flavor $U(1)$ generators that consist with the global symmetry group $\mathcal{F}$. Then by using the expression for the central charge written in terms of 't Hooft anomalies: 
$$a_{trial}(s)=\frac{3}{32}(3TrR_{t}^{3}-TrR_{t}) $$
the problem reduced to fining the right values of $s_{I}$ which corresponds to the maximization of the central charge:
$$\frac{\partial a_{trial}}{\partial s_{I}}=\frac{3}{32}(9TrR_{t}^{2}F_{I}-TrF_{I})=0 $$
or
$$9TrR_{t}^{2}F_{I}=TrF_{I} $$
while to ensure local maximum we require that the matrix of second derivatives
$$\frac{\partial^{2}a_{trial}}{\partial s_{I}\partial s_{J}}=\frac{27}{16}TrR_{trial}F_{I}F_{J}<0 $$
is negative-definite. As IW shown in [8] the last two equations are always true for any unitary superconformal field theory.
\newline In the following sections we will apply the last procedure of finding the R-symmetry to cases of SQCD with one or two adjoint field. Through the process we will fix our central charge each time the unitarity bound is violated, that will schematically the process looks like:
\begin{figure}
\begin{center}
\includegraphics[scale=0.44]{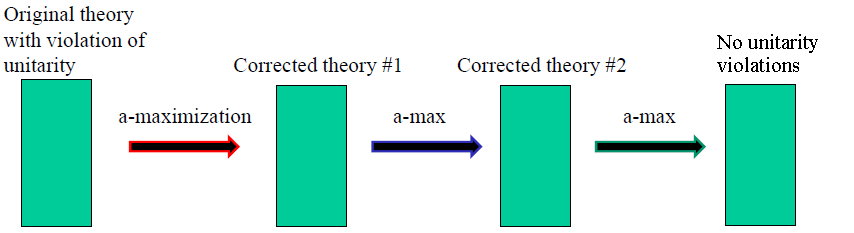}
\end{center}
\caption{\footnotesize\textit{Each a-maximization procedure changes the R charges}}
\end{figure}
The way to do that is explained in [5]:
$$a\rightarrow\widetilde{a}=a+\sum_{M}\left[a_{M}\left(\frac{2}{3}\right)-a_{M}(R(M))\right]= $$
$$a+\frac{1}{9}\sum_{M}dim(M)[2-3R(M)]^{2}[5-3R(M)] $$
where $M$ is the set of all the gauge invariant chiral superfields whose R charge is less than $\frac{2}{3}$.
The relevant part in our case is to find the set $M$ thereby estimating the decoupled fields in the IR and estimating $f_{IR}$ from below (then by comparing that to $f_{UV}$ the ACS conjecture can be checked). Of course, the major weakness of the above method is that true value of $f_{IR}$ will often be larger due to interaction contributions. Nevertheless, these estimations will be non-trivial in many examples as will be shown in the next sections.\newline
The theories that we will handle first have $SU(N_{c})$ gauge group and include just one adjoint $X$, then we move on to include two adjoint fields $X$ and $Y$ which is the maximal number of adjoint fields compatible with asymptotic freedom [9].
The possible RG fixed points along with their corresponding superpotential deformation can be classified as follows:
\begin{center}
\includegraphics[scale=0.78]{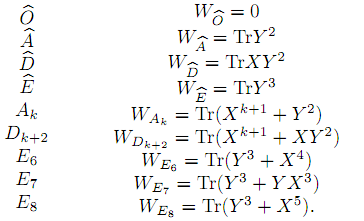}
\end{center}
It turns out that they sit on the same structure as appears in Arnold's theory of singularities and known as $ADE$ classification. As we see from above the simplest way is to start with $\hat{O} $ type deformation where $W=0$. What are the possible deformations that could drive RG flows from this point?\newline
Let us consider superpotentials involving only the adjoints, $W=TrX^{k}Y^{l} $. We can show that $R(X)=R(Y)>\frac{1}{2} $ independent of $N_{c}$ and $N_{f}$ values thus $W$ is relevant deformation of $\hat{O}$ only when $k+l\leq3 $. Using the last argument to find the relevant terms for RG flows and reveal the full structure of superpotential deformations starting from $\hat{O}$ which appears in the figure 2 below.
\begin{figure}
\begin{center}
\qquad\qquad\includegraphics[scale=0.75]{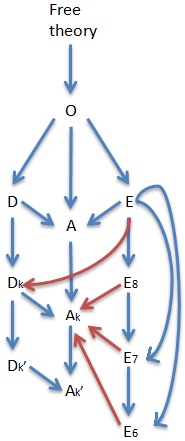}
\end{center}
\caption{\footnotesize\textit{ADE constuction of possible RG flows. Red lines indicates indicate flow to a particular value of $k$.}}
\end{figure}
\newpage
\subsection{Tests on SQCD with one adjoint}
In adjoint SQCD we have three different types of gauge invariant chiral superfields that are relevant for the following discussion:\newline\newline
a. The Baryon: $\mathcal{B}^{(n_{1},n_{2},...,n_{k})}=Q_{(1)}^{n_{1}}Q_{(2)}^{n_{2}}\cdots Q_{(k)}^{n_{k}}$, $\sum_{l=1}^{k}n_{l}=N_{c}$  $n_{l}\leq N_{f}$ for $k\in\mathbb{Z}$. 
In the last expression the color indices are contracted with an $\epsilon$ tensor and we introduced the notation of $Q_{(l)}=X^{l-1}Q$. We also kept in mind that also anti-baryon $\widetilde{\mathcal{B}} $ exist which can be expressed by making the change $Q\rightarrow\widetilde{Q}$ in the expression for $\mathcal{B}$.
Denote $R(Q)=y$ then $R(X)=\frac{1-y}{x}$ and we can write:
$$R(\mathcal{B}^{(n_{1},n_{2},...,n_{k})})=\sum_{l=1}^{k}n_{l}(l-1)\frac{1-y}{N_{c}}N_{f}+N_{c}y$$
b. The Meson: $(\mathcal{M}_{j})_{\widetilde{i}}^{i}=\widetilde{Q}_{\widetilde{i}}X^{j-1}Q^{i}$ for $j\in\mathbb{Z}$. \newline
$$R(\mathcal{M}_{j})=2y+(j-1)\frac{1-y}{N_{c}}N_{f}$$
c. $trX^{j-1}$ for $j\in\mathbb{Z}-\{1\}$
$$R(trX^{j-1})=(j-1)\frac{1-y}{N_{c}}N_{f}$$
\newline\newline
In order to determine the central charge we introduce an auxiliary quantity, the trial central charge computed with the assumption that the first $m$ operators of type one are free, $n$ operators of type two are free, $p$ operators of type three are free:
\begin{equation}\begin{split}&\widetilde{a}^{(m,n,p)}=6N_{c}N_{f}(y-1)^{3}-2N_{c}N_{f}(y-1)+3(N_{c}^{2}-1)\left[\frac{1-y}{N_{c}}N_{f}-1\right]^{3}-(N_{c}^{2}-1)\left[\frac{1-y}{N_{c}}N_{f}-1\right]+\\
&2N_{c}^{2}+\frac{1}{9}\sum_{j=1}^{m}\left[2-3\left\{ 2y+(j-1)\frac{1-y}{N_{c}}N_{f}\right\} \right]^{2}\left[5-3\left\{ 2y+(j-1)\frac{1-y}{N_{c}}N_{f}\right\} \right]\\
&+\frac{1}{9}\sum_{j=2}^{n}\left[2-3\left\{ (j-1)\frac{1-y}{N_{c}}N_{f}\right\} \right]^{2}\left[5-3\left\{ (j-1)\frac{1-y}{N_{c}}N_{f}\right\} \right]\\
&+\frac{1}{9}\sum_{j=1}^{p}\left[2-3\left\{ \sum_{l=1}^{j}n_{l}(l-1)\frac{1-y}{N_{c}}N_{f}+N_{c}(y-1)\right\} \right]^{2}\left[5-3\left\{ \sum_{l=1}^{j}n_{l}(l-1)\frac{1-y}{N_{c}}N_{f}+N_{c}(y-1)\right\} \right]
\end{split}\end{equation}
We would like to determine how many of chiral superfields have dimension of less than $\frac{2}{3}$. For simplification we will work here the case where $N_{c}=2$ and $N_{f}\in{1,2,3}$, even that the same procedure can be applied for higher values.\newline\newline
The method in which we'll work here is via iterative algorithm - \newline
a. Starting first with $a^{(0,0,0)}$, determining its maximization argument \newline
b. Checking whether an operator of the three possible become free.\newline
c. Updating the trial central charge accordingly. (e.g after one step we arrive at $a^{(1,0,0)}$ or $a^{(0,1,0)}$ or $a^{(0,0,1)}$).\newline
d. The iteration continues as long as we can decouple one of the possible fields.\newline
\newline
Inserting the above information to mathematica we arrive at the following values:
$$y_{max}^{(0)}\mid_{N_{f}=1}=\frac{105-2\sqrt{1009}}{87}\approx0.47667 $$
$$y_{max}^{(0)}\mid_{N_{f}=2}=\frac{24-2\sqrt{241}}{15}\approx0.56505$$
$$y_{max}^{(0)}\mid_{N_{f}=3}=\frac{69-2\sqrt{889}}{15}\approx0.62452$$\newline\newline
After the first iteration of the first case only the chiral superfield $trX^{2}$ became free (with $R(trX^{2})\simeq0.52333)$.\newline
Maximizing $a^{(0,0,1)}$ we find $y_{max}^{(1)}\mid_{N_{f}=1}=\frac{17-2\sqrt{21}}{15}\approx0.52232$. Checking the three possible operators we find that each of them are now having $R$ charge of more than $\frac{2}{3}$.
Thus in this case only one chiral operator became free.\newline\newline
In the case where $N_{f}=2$ we find that $y_{max}^{(0)}\mid_{N_{f}=2}\approx0.56506 $ leads to $R(X)\approx0.43494$ thus no invariant will violate the unitarity law as $R(trX^{2})>\frac{2}{3}$. Similarly in the case of $N_{f}=3$ we find that all the chiral operators have dimension of more than $\frac{2}{3}$ thus all of them still confined. \newline
The expression for $f_{UV}$ obtained by assuming asymptotic freedom is:
$$f_{UV}=2(N_{c}^{2}-1)+4N_{f}N_{c}$$
While the expression for $f_{IR}$ is:
$$f_{IR}=8$$
$$f_{IR}=0$$
$$f_{IR}=0$$
for $N_{f}\in\{1,2,3\}$ respectively. It is not hard to see that in these cases ACS is easily vindicated in all cases.

\subsubsection{Tests with large number of $N_{c}$ and $N_{f}$}
We would now check the case where both $N_{c}$ and $N_{f}$ are very large and also $N_{f}\ll N_{c}$. Assuming $R(Q)$ is still finite in the last case, the trial central charge becomes:
$$\lim_{x\rightarrow\infty}a^{(0)}=6N_{f}^{2}x(y-1)^{3}-10N_{f}^{2}x(y-1) $$
where $x\equiv\frac{N_{c}}{N_{f}}$. As explained before, we assume a decoupling of operators that violate the unitarity bound. In order to ``correct" the trial charge in our limit we pass to an integral by replacing the sum over $j$ by integral over $t$ which is defined by:
$$ t=2y+(j-1)\frac{1-y}{x} $$
Noting that in the $x\rightarrow\infty $ limit only the meson operators $\mathcal{M}_{j}$ are relevant and performing the integration yields:
$$\frac{1}{9}\sum_{j=1}^{p}\left[2-3\left(2y+(j-1)\frac{1-y}{x}\right)\right]^{2}\left[5-3\left(2y+(j-1)\frac{1-y}{x}\right)\right]\rightarrow\frac{N_{f}^{2}}{18}x(2-6y)^{3} $$
and our trial charge becomes:
$$a=6N_{f}^{2}x(y-1)^{3}-10N_{f}^{2}x(y-1)+\frac{N_{f}^{2}}{18}x(2-6y)^{3} $$
The argument corresponds to its local maximum is
$$y_{max}=\frac{\sqrt{3}-1}{3} $$
We can now estimate the number of the decoupled fields by extracing $p$ from the equation
$$2y+(p-1)\frac{1-y}{x}=\frac{2}{3} $$
Which yields $p\approx\frac{10-4\sqrt{3}}{13}x=\frac{10-4\sqrt{3}}{13}\frac{N_{c}}{N_{f}}$.\newline
Thus $f_{UV}$ and $f_{IR}$ can be estimated as:
$$f_{UV}= 2(N_{c}^{2}-1)+4N_{c}N_{f}$$
$$\ensuremath{f_{IR}}=-\lim_{T\rightarrow 0}\frac{90f(T)}{T^{4}\pi^{2}}\geq 2\frac{10-4\sqrt{3}}{13}N_{c}N_{f}$$
\newline And we can see that ACS prediction is indeed verified as:
$$2\frac{10-4\sqrt{3}}{13}N_{c}N_{f}\leq2(N_{c}^{2}-1) + 4N_{c}N_{f} \quad \rightarrow \quad \frac{10-4\sqrt{3}}{13}<x+2 $$

\subsection{Tests on SQCD with two adjoints}
The case studied in the last two sections can be made more general by working with a larger set of particles in the adjoint. Unfortunatelly, only a small subset of these theories are asymptotically free as $N_{a}=2$ is the maximum
number of particles that compatible with asymptotic freedom. \newline\newline
We will start by giving the chiral ring of operators, then analyze the family of SQCD with two adjoints $X$ and $Y$ in case of $N_{c}=2$ and $N_{f}\in\{1,2,3\}$ .\newline

\subsubsection{Chiral ring of operators}
There are three possible operators (which are generalization of the previous case):\newline
a. $\mathcal{O}_{I_{1},...,I_{n}}=TrX_{I_{1}}...X_{I_{n}}$\newline 
b. $(\mathcal{M}_{I_{1},...,I_{n}})_{i\widetilde{i}}=\widetilde{Q}_{\widetilde{i}}X_{I_{1}}...X_{I_{n}}Q_{i} $\newline
c. $\mathcal{B}^{(n_{(I_{1},...,I_{n_{1}})},n_{(J_{1},...,J_{n_{2}})},...,n_{(K_{1},...,K_{n_{1}})})}=Q_{(I_{1},...,I_{n_{1}})}^{n_{(I_{1},...,I_{n_{1}})}}Q_{(J_{1},...,J_{n_{1}})}^{n_{(J_{1},...,J_{n_{2}})}}...Q_{(K_{1},...,K_{n_{1}})}^{n_{(K_{1},...,K_{n_{1}})}} $\newline
where $N_{c}=n_{(I_{1},...,I_{n_{1}})} + ... + n_{(K_{1},...,K{}_{n_{k}})} $ while $n_{(I_{1},...,I_{n_{1}})}\leq N_{f} $ and the dressed quarks $Q_{(I_{1},...,I_{n})}=X_{I_{1}},...,X_{I_{n}}Q $ are fully contracted with an epsilon tensor.\newline
We will denote the anomaly-free trial R-symmetry as:\newline
$$R(Q_{i})=R(\widetilde{Q}_{i})\equiv y,$$ $$R(Y)\equiv z,$$ $$R(X)\equiv1-z+\frac{1-y}{x} $$
Rather than writing the cumbersome expressions for the corresponding R-symmetries of the various possible chiral operators in the general case we will explicitly write only the case of two possible types of adjoint chiral operators.\newline
Denote $I\in\{0,1\}  $ such that $X_{0}\equiv X $ and $X_{1}\equiv Y $ and $\sum_{j=1}^{j=n}I_{j}\equiv m_{I}$ where $I$ denotes any possible set with number of elements equal to $n_{I}$ then:
$$R(\mathcal{O}_{I_{1},...,I_{n}})=m_{I}z + (n-m_{I})\left(1-z + \frac{1-y}{N_{c}}N_{f}\right),$$
$$R(\mathcal{M}_{I_{1},...,I_{n}})=2y + m_{I}z+(n-m_{I})\left(1-z + \frac{1-y}{N_{c}}N_{f}\right),$$
$$R(\mathcal{B}^{(n_{(I_{1}...I_{n_{1}})},n_{(J_{1}...J_{n_{2}})},...,n_{(K_{1}...K_{n_{1}})})})=yN_{c}+\sum n_{(I_{1},...,I_{n_{1}})}\left\{ m_{I}z+(n_{I}-m_{I})\left(1-z+\frac{1-y}{x}\right)\right\}  $$
\subsubsection{The $\hat{O}$ RG fixed points $W_{\hat{O}}=0$}
The trial central charge can be written as:
\begin{equation}\begin{split}&\widetilde{a}^{(m,n,p)}=6N_{c}N_{f}(y-1)^{3}-2N_{c}N_{f}(y-1)+3(N_{c}^{2}-1)\left[\frac{1-y}{N_{c}}N_{f}-z\right]^{3}-(N_{c}^{2}-1)\left[\frac{1-y}{N_{c}}N_{f}-z\right]\\
&+3(N_{c}^{2}-1)(z-1)^{3}-(N_{c}^{2}-1)(z-1)+2N_{c}^{2}+\frac{1}{9}\sum_{j=1}^{m}[2-3R(\mathcal{O})]^{2}\left[5-3R(\mathcal{O}) \right]\\
&+\frac{1}{9}\sum_{j=1}^{n}\left[2-3R(\mathcal{M}) \right]^{2}\left[5-3R(\mathcal{M}) \right]+\frac{1}{9}\sum_{j=1}^{p}\left[2-3R(\mathcal{B}) \right]^{2}\left[5-3R(\mathcal{B}) \right]
\end{split}\end{equation}
The arguments corresponding to its maximal value are:
$$(R^{(0)}(Q),R^{(0)}(X),R^{(0)}(Y))\mid_{N_{f}=1}\approx(0.63639, 0.5909, 0.5909)  $$

$$(R^{(0)}(Q),R^{(0)}(X),R^{(0)}(Y))\mid_{N_{f}=2}\approx(0.66667, 0.66667, 0.66667) $$

$$(R^{(0)}(Q),R^{(0)}(X),R^{(0)}(Y))\mid_{N_{f}=3}\approx(0.68977, 0.73267, 0.73267) $$\newline
Any of the possible chiral operators in the IR will respect unitarity, thus there is no need for further corrections in this case. As there is no decoupled fields in the IR the check for ACS in this case is obvious and is similar to what we have done in the last part.

\subsubsection{The $\hat{E}$ RG fixed points $W_{\hat{E}}=Y^{3}$}
Denote the $R$-charges as follows:
$$ R(Q_{i})=R(\widetilde{Q}_{i})\equiv y,$$ $$R(Y)=\frac{2}{3},$$ $$R(X)\equiv\frac{1-y}{x}+\frac{1}{3} $$
\begin{equation}\begin{split}&\widetilde{a}^{(m,n,p)}=6N_{c}N_{f}(y-1)^{3}-2N_{c}N_{f}(y-1)+3(N_{c}^{2}-1)\left[\frac{1-y}{x}-\frac{2}{3}\right]^{3}-(N_{c}^{2}-1)\left[\frac{1-y}{x}-\frac{2}{3}\right]\\
&+3(N_{c}^{2}-1)\left[\frac{2}{3}-1\right]^{3}-(N_{c}^{2}-1)\left[\frac{2}{3}-1\right]+2N_{c}^{2}+\frac{1}{9}\sum_{j=1}^{m}[2-3R(\mathcal{O})]^{2}\left[5-3R(\mathcal{O})\right] \\
&+\frac{1}{9}\sum_{j=1}^{n}\left[2-3R(\mathcal{M}) \right]^{2}\left[5-3R(\mathcal{M}) \right]+\frac{1}{9}\sum_{j=1}^{p}\left[2-3R(\mathcal{B}) \right]^{2}\left[5-3R(\mathcal{B}) \right]
\end{split}\end{equation}
The arguments corresponding to its maximal value are:
$$(R^{(0)}(Q), R^{(0)}(X), R^{(0)}(Y))\mid_{N_{f}=1}\approx(0.60919, 0.91954, 0.66666)  $$

$$(R^{(0)}(Q), R^{(0)}(X),R^{(0)}(Y))\mid_{N_{f}=2}\approx(0.66666, 0.83333, 0.66666) $$

$$(R^{(0)}(Q), R^{(0)}(X), R^{(0)}(Y))\mid_{N_{f}=3}\approx(0.7035, 0.77808, 0.66666) $$\newline
\subsubsection{The $\hat{D}$ RG fixed points $W_{\hat{D}}=XY^{2}$}
Denote the $R$-charges as follows:
$$ R(Q_{i})=R(\widetilde{Q}_{i})\equiv y, $$ $$R(Y)\equiv\frac{y-1}{x}+1,$$ $$R(X)\equiv\frac{2-2y}{x}$$

\begin{equation}\begin{split}&\widetilde{a}^{(m,n,p)}=6N_{c}N_{f}(y-1)^{3}-2N_{c}N_{f}(y-1)+3(N_{c}^{2}-1)\left[\frac{y-1}{x}\right]^{3}-(N_{c}^{2}-1)\left[\frac{y-1}{x}\right]\\
&+3(N_{c}^{2}-1)\left[\frac{2-2y}{x}-1\right]^{3}-(N_{c}^{2}-1)\left[\frac{2-2y}{x}-1\right]+2N_{c}^{2}+\frac{1}{9}\sum_{j=1}^{m}[2-3R(\mathcal{O})]^{2}\left[5-3R(\mathcal{O})\right] \\
&+\frac{1}{9}\sum_{j=1}^{n}\left[2-3R(\mathcal{M}) \right]^{2}\left[5-3R(\mathcal{M}) \right]+\frac{1}{9}\sum_{j=1}^{p}\left[2-3R(\mathcal{B}) \right]^{2}\left[5-3R(\mathcal{B}) \right]
\end{split}\end{equation}
The arguments corresponding to its maximal value are:
$$(R^{(0)}(Q), R^{(0)}(X), R^{(0)}(Y))\mid_{N_{f}=1}\approx(0.62837, 0.37163, 0.81419)  $$

$$(R^{(0)}(Q), R^{(0)}(X),R^{(0)}(Y))\mid_{N_{f}=2}\approx(0.20474, 1.59052, 0.20474) $$

$$(R^{(0)}(Q), R^{(0)}(X), R^{(0)}(Y))\mid_{N_{f}=3}\approx(0.63115, 1.10655, 0.44673) $$\newline
Thus only in the second case there's violation of unitarity. The first operators to decouple will be the meson $\mathcal{M}=\widetilde{Q}_{\widetilde{i}}Q_{i} $ and $\mathcal{O}_{1,1}=Tr Y^{2}$ and . Maximizing $a^{(0,1,0)}$ we find:
$$(R^{(1)}(Q),R^{(1)}(X),R^{(1)}(Y))\mid_{N_{f}=2}\approx(0.26697, 1.46606, 0.26697)  $$
Maximizing $a^{(1,1,0)}$ we find:
$$(R^{(2)}(Q),R^{(2)}(X),R^{(2)}(Y))\mid_{N_{f}=2}\approx(0.32115, 1.3577, 0.32115)  $$
Next operators to be decouple is either $\mathcal{M}=\widetilde{Q}_{\widetilde{i}}YQ_{i} $ or $\mathcal{O}_{1,1,1}=Tr Y^{3}$, here both have $R$ charge higher than $\frac{2}{3}$ thus no further corrections to the central charge is needed. Let us check $f_{UV}$ and $f_{IR}$:
$$f_{UV}=2(N_{c}^{2}-1)+4N_{f}N_{c}=22$$
$$f_{IR}=8+8=16$$
ACS it still vindicated.

\subsection{Analysis of chiral theories}
\subsubsection{Deconfinment and mixed phase}
Up until now our checks in this chapter involves $SU(N)$ gauge theory with one or two adjoints and various superpotential deformations, all of which are vector-like theories. However, some of the most interesting SUSY gauge theories are chiral theories, since these experience dynamical supersymmetry breaking.
The simplest models of supersymmetry breaking usually making use of an antisymmetric tensor and some number of flavors. The dynamics of these theories is well-understood only in case of small number of flavors, however one could not conclusively find the low-energy phase of such a theory in the general case.\newline
We now would like to follow [19] and check ACS on a $SU(N_{c})$ gauge theory with a two-index anti-symmetric tensor and $N_{f}$ flavors in the fundamental while $N_{c}+N_{f}-4$ in the anti-fundamental and no tree-level superpotential, all these transformation properties are summarized below:
\begin{center}
\includegraphics[scale=0.8]{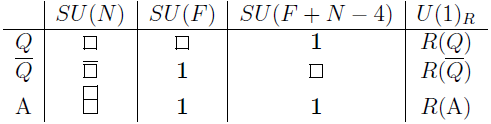}
\end{center}
The values of the R-charges which results from the vanishing of the NSVZ $\beta$ function are:
\begin{equation}R(A)=\frac{2N_{f}-6-N_{f}R(Q)-(N_{f}+N_{c}-4)R(\widetilde{Q})}{N_{c}-2}\end{equation}
After denoting $R(Q)\equiv y $ and $R(\bar{Q})\equiv z $ we can write the trial central charge as:
\begin{equation}\begin{split}&\widetilde{a}^{(0)}=3N_{c}N_{f}(y-1)^{3}-N_{c}N_{f}(y-1)+3N_{c}(N_{f}+N_{c}-4)(z-1)^{3}-N_{c}(N_{f}+N_{c}-4)(z-1)+\\
&3\frac{N_{c}(N_{c}-1)}{2}\left[\frac{2N_{f}-6-N_{f}y-(N_{f}+N_{c}-4)z}{N_{c}-2}-1\right]^{3}-\\
&\frac{N_{c}(N_{c}-1)}{2}\left[\frac{2N_{f}-6-N_{f}y-(N_{f}+N_{c}-4)z}{N_{c}-2}-1\right]
\end{split}\end{equation}
The chiral ring of operators are consists of two types:\newline
a. Meson - $\mathcal{M}=Q\bar{Q}$ and $H=\bar{Q}A\bar{Q}$.\newline
b. Baryons - $B_{k}=Q^{k}A^{\frac{N_{c}-k}{2}} $ for $k$ and $N$ both even or odd and $k\leq min(N_{c},N_{f})$, and also $B=\bar{Q}^{N_{c}}$.\newline
The tests we would like to preform here involves $N_{c}=5$ and $N_{f}\in(5,6,7)$. The arguments corresponding to the maximal value of the central charge are:
$$(R^{(0)}(Q), R^{(0)}(\bar{Q}),R^{(0)}(A))\mid_{N_{f}=5}=(0.31166, 0.31166, 0.22502)$$
$$(R^{(0)}(Q), R^{(0)}(\bar{Q}),R^{(0)}(A))\mid_{N_{f}=6}=(0.39337, 0.39337, 0.29539)$$
$$(R^{(0)}(Q), R^{(0)}(\bar{Q}),R^{(0)}(A))\mid_{N_{f}=7}=(0.45756, 0.45756, 0.37887)$$
We note that the R-charges of $Q$ and $\bar{Q}$ are the same even though the theory is chiral. This is a consequence of the absence of superpotential as gauge interactions cannot distinguish between the two fields. This property will be violated as we introduce the corrections to the central charge.
\newline The only case in which we have unitarity violation is when $N_{f}=5$ where the first smallest invariant to become free is the $Q\bar{Q}$ meson.
After the first iteration the correction to the central charge will be:
\begin{equation}\begin{split}&a^{(0)}+\frac{1}{9}dim(M)(2-3R(M))^{2}(5-3R(M))=\\
&a^{(0)}+\frac{N_{f}}{9}(N_{f}+N_{c}-4)\left(2-3R(Q)-3R(\bar{Q})\right)^{2}\left(5-3R(Q)-3R(\bar{Q})\right).
\end{split}\end{equation}
as $dim(M)=N_{f}(N_{f}+N_{c}-4)$. Correcting the central charge as described above and maximizing the above $a^{(1)}$ we find:
$$(R^{(1)}(Q), R^{(1)}(\bar{Q}), R^{(1)}(A))\mid_{N_{f}=5}=(0.29853, 0.30512, 0.22554)$$
As $R(H)>\frac{2}{3} $ no further gauge invariant will become free. The general expression for the free energy in the $UV$ is:
$$f_{UV}=2(N_{c}^{2}-1)+2N_{f}N_{c}+2N_{c}(N_{f}+N_{c}-4) $$
For the last two cases we can immediately say that ACS is vindicated since no gauge invariant became free while for $N_{f}=5$ we have:
$$f_{UV}\mid_{N_{f}=5}=142$$
compared to the free energy in the IR:
$$f_{IR}\mid_{N_{f}=5}=60$$
and we can see that ACS is vindicated again.

Up to now all the theories we analyzed were either non-chiral or chiral with no superpotential which results in the same $R$-charge for both $Q$ and $\widetilde{Q}$. This assumption will no longer holds for the next two theories we analyze here as they are chiral theories with non-vanishing superpotential.
\subsubsection{Self-dual chiral theory}
Following [26], the matter content will be given by:
\begin{center}
\includegraphics[scale=0.75]{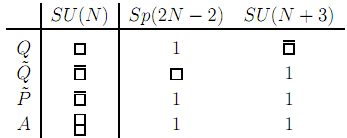}
\end{center}
with superpotential $W=\widetilde{Q}A\widetilde{Q}$ (which is the most general renormalizable form of superpotential compatible with the symmetries). As there is no mention of flavor in this theory we shall use just $N$ instead of the usual $N_{c}$.\newline
Similarly to previous cases we have two types of particles:\newline
1. Mesons - $Q\widetilde{Q}$, $Q\widetilde{P}$, $\widetilde{Q}A\widetilde{Q}$ and $\widetilde{Q}A\widetilde{P}$.\newline
2. Baryons - $\widetilde{Q}^{N} $, $\widetilde{Q}^{N-1}\widetilde{P}$ and $Q^{k}A^{\frac{(N-k)}{2}}$ for $k=1, 3, ..., N$.\newline\newline
Let us now relate both $R$-charge for $A$ and $\widetilde{P}$ to those of $Q$ and $\widetilde{Q}$. In order to do that we use two constraints - the fist comes from the superpotential dimension while the second from the ABJ anomaly cancellation:
\begin{equation}R(A)=-2(R(\widetilde{Q})-1)\end{equation}
\begin{equation}T(G)+\sum_{i}T(r_{i})(R_{i}-1)=0 \end{equation}
\begin{equation}N+\frac{N+3}{2}(R(Q)-1)+\frac{2N-2}{2}(R(\widetilde{Q})-1)+\frac{1}{2}(R(\widetilde{P})-1)+\frac{N-2}{2}(R(A)-1)=0 \end{equation}
\begin{equation}R(\widetilde{P})=4-(N+3)R(Q)-2R(\widetilde{Q})\end{equation}
Using the last expressions we can write the central charge $a$ as two parameters function:
\begin{equation}\begin{split}&a^{(0)}=2(N^{2}-1)+N(N+3)\left(3(R(Q)-1)^{3}-(R(Q)-1)\right)+N(2N-2)(3(R(\widetilde{Q})-1)^{3}-\\
&(R(\widetilde{Q})-1))+N\left(3(R(\widetilde{P})-1)^{3}-(R(\widetilde{P})-1)\right)+\frac{N}{2}(N-1)\left(3(R(A)-1)^{3}-(R(A)-1)\right) 
\end{split}\end{equation}
Substituting $(9.8)$ and $(9.11)$ in the central charge expression and requiring that
\begin{equation}\frac{\partial a}{\partial R(Q)}=\frac{\partial a}{\partial R(\widetilde{Q})}=0 \end{equation}
we find:
\begin{equation} R(\widetilde{Q})=\frac{1}{2}\left(4-(4+N)R(Q)\right) \end{equation}
and the explicit form for $R(Q)$ in terms of $N$ is:
\begin{equation} R(Q)=\frac{12-12N-4N^{2}+\frac{4}{3}\sqrt{N^{4}+4N^{3}+5N^{2}-18N+9}}{12-8N-7N^{2}-N^{3}} \end{equation}
Now let us examine how many invariants from the chiral ring will become free for theories with $N=2, 3, 4$ the argument corresponding to the central charge maximum are:
$$(R^{(0)}(Q),R^{(0)}(\widetilde{Q}),R^{(0)}(\widetilde{P}), R^{(0)}(A))\mid_{N=2}=(0.50851, 0.47445, 0.50851, 1.05110) $$
$$(R^{(0)}(Q),R^{(0)}(\widetilde{Q}),R^{(0)}(\widetilde{P}), R^{(0)}(A))\mid_{N=3}=(0.40852, 0.57016, 0.40852, 0.85968) $$
$$(R^{(0)}(Q),R^{(0)}(\widetilde{Q}),R^{(0)}(\widetilde{P}), R^{(0)}(A))\mid_{N=4}=(0.35374, 0.58503, 0.35374, 0.82994) $$
$$(R^{(0)}(Q),R^{(0)}(\widetilde{Q}),R^{(0)}(\widetilde{P}), R^{(0)}(A))\mid_{N=5}=(0.31223, 0.59495, 0.31223, 0.81011) $$
In first four cases there is no invariant that become free thus the ACS conjecture is automatically vindicated. A much interesting case from ACS point of view is $N=5$ where the invariant $Q\widetilde{P}$ become free. 
\newline As $dim(Q\widetilde{P})=8$ and $R(Q\widetilde{P})=R(Q)+R(\widetilde{P})=4-7R(Q)-2R(\widetilde{Q})$ the correction to the central charge is:\newline
\begin{equation}a^{(1)}=a^{(0)}+\frac{8}{9}(21R(Q)+6R(\widetilde{Q})-10)^{2}(21R(Q)+6R(\widetilde{Q})-10)\end{equation}
Second iteration yields:
$$(R^{(1)}(Q),R^{(1)}(\widetilde{Q}),R^{(1)}(\widetilde{P}, R^{(0)}(A)))\mid_{N=5}=(0.31242, 0.59509, 0.31047, 0.80982)$$
As this is the only invariant the will become free we can estimate the free energy in the $IR$ as $f_{IR}=8$ which is lower than $f_{UV}=2N(N+3)+2N(2N-2)+2N+N(N-1)=190$.
\subsubsection{Three types of flavors}
We adopt here the same notation for flavors as in [27]. The matter content will be given by:
\begin{center}
\includegraphics[scale=0.75]{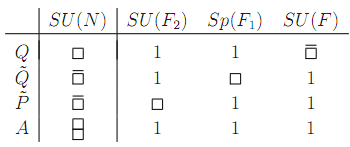}
\end{center}
with superpotential $W=\widetilde{Q}A\widetilde{Q}$.
The chiral ring of operators is:\newline
1. Meson - $\widetilde{P}Q$, $\widetilde{Q}Q$ and $\widetilde{P}A\widetilde{P}$.\newline
2. Baryon - $\widetilde{P}^{N} $, for odd $N$ $QA^{\frac{N-1}{2}}, Q^{3}A^{\frac{N-3}{2}},..., Q^{k}A^{\frac{N-k}{2}}$ while for even $N$ $A^{\frac{N}{2}}, Q^{2}A^{\frac{N-2}{2}},..., \\Q^{k}A^{\frac{N-k}{2}}$ where $k\leq min(N,F)$.\newline
Using the constraints we can relate the $R$-charges according to:\newline
\begin{equation}R(A)=-2(R(\widetilde{Q})-1)\end{equation}
\begin{equation}R(\widetilde{P})=\frac{F+F_{1}+F_{2}-3N+2-FR(Q)+(2N-F_{1}-4)R(\widetilde{Q})}{F_{2}} \end{equation}
\begin{equation}F_{2}=N+F-F_{1}-4 \end{equation}
We can write down the central charge of this theory as:
\begin{equation}\begin{split}&a^{(0)}=NF\left(3(R(Q)-1)^{3}-(R(Q)-1)\right)+NF_{1}\left(3(R(\widetilde{Q})-1)^{3}-(R(\widetilde{Q})-1)\right)+\\
&NF_{2}\left(3(R(\widetilde{P})-1)^{3}-(R(\widetilde{P})-1)\right)+\frac{N}{2}(N-1)\left(3(R(A)-1)^{3}-(R(A)-1)\right) 
\end{split}\end{equation}
The theories that we would like to check ACS will be $(N,F,F_{1},F_{2})=(7,7,2,8), (8,7,3,8)$ and $(9,7,4,8)$. Eliminating $R(A)$ and $R(P)$ using the above constraints and looking for its maximization as in $(9.11)$ we find:
$$(R(Q),R(\widetilde{Q}),R(\widetilde{P}),R(A))\mid_{(N,F,F_{1},F_{2})=(7,7,2,8)}=(0.32816, 0.86531, 0.32816, 0.26938)$$
$$(R(Q),R(\widetilde{Q}),R(\widetilde{P}),R(A))\mid_{(N,F,F_{1},F_{2})=(8,7,3,8)}=(0.26838, 0.89175, 0.26838, 0.21650)$$
$$(R(Q),R(\widetilde{Q}),R(\widetilde{P}),R(A))\mid_{(N,F,F_{1},F_{2})=(9,7,4,8)}=(0.21162, 0.91743, 0.21162, 0.16514)$$
First particle to become free is $\widetilde{P}Q$, after first iteration:
$$(R(Q),R(\widetilde{Q}),R(\widetilde{P}),R(A))\mid_{(N,F,F_{1},F_{2})=(7,7,2,8)}=(0.32578, 0.86186, 0.32679, 0.27628)$$
$$(R(Q),R(\widetilde{Q}),R(\widetilde{P}),R(A))\mid_{(N,F,F_{1},F_{2})=(8,7,3,8)}=(0.23472, 0.84712, 0.24764, 0.30576)$$
$$(R(Q),R(\widetilde{Q}),R(\widetilde{P}),R(A))\mid_{(N,F,F_{1},F_{2})=(9,7,4,8)}=(0.14225, 0.83266, 0.16636, 0.33468)$$
No further invariants will become free as all the next invariants obey the unitarity law. Let us now estimate the free energy both in the $IR$ and the $UV$:
$$f_{UV}=2(N^{2}-1)+2NF+2NF_{1}+2NF_{2}+N(N-1)$$
$$f_{IR}=2dim(PQ)=2FF_{2}$$
which confirms with ACS in all cases as:
$$f_{UV}=376 > f_{IR}=112$$
$$f_{UV}=470 > f_{IR}=112$$
$$f_{UV}=574 > f_{IR}=112$$
in the respective cases.
\section{Conclusions}
All along this paper we passed over eleven different tests to check the power of the ACS conjecture with all of them yielding the same result (that the ACS conjecture is vindicated). Duality and a-maximization have played a cucial role along this work and paved us the way to extract information about the IR region which cannot be done otherwise nowadays.
The bottom line is that a convincing pieces of evidence have been gathered to support the hope that a proof for the ACS conjecture can be found.
\\
\\
\\
\\
\\
\\
\\
\\
\begin{center}
 \bf{Acknowledgement}
\end{center}
Y.M. would like to thank Zohar Komargodski for suggesting this project.
\newpage

\textbf{Appendix A. Tests in Infrared-Free Supersymmetric Examples}\\\\
In this section we will reproduce the results that are already known in the original paper [1]. All the cases discussed here will have an infrared-free dual which means that we can completely discard correction terms above $O(g^{0})$.
Ofcourse such infrared-free duals will exist only under appropriate conditions on the number of colors and flavors, but the interesting part of these result is that ACS inequality is saturated exactly at these points.
\\\\\textit{A.1. $SU(N_{c})$ gauge group}\\\\
Using the expressions $(3.10)$ and $(3.11)$, we find that:
\begin{equation}f_{UV}=[2(N_{c}^{2}-1)+4N_{c}N_{f}]\left(1+\frac{7}{8}\right)\end{equation}
\begin{equation}f_{IR}=[2((N_{f}-N_{c})^{2}-1)+4(N_{f}-N_{c})N_{f}+2N_{f}^{2}]\left(1+\frac{7}{8}\right)\end{equation}
and the ACS inequality becomes:
\begin{equation}2((N_{f}-N_{c})^{2}-1)+4(N_{f}-N_{c})N_{f}+2N_{f}^{2}\leq2(N_{c}^{2}-1)+4N_{c}N_{f}\end{equation}
\begin{equation}N_{f}\leq\frac{3}{2}N_{c}\end{equation}
Remarkably, we get the condition which is corresponding precisely to the boundary of the weak magnetic phase determined by the analysis of Seiberg. The inequality will hold also for $N_{f}>\frac{3}{2}N_{c}$ but this time the dual theory will not be infrared dual and higher orders of the free energy should be taken into account (see section 2).
\\\\\textit{A.2. $SO(N_{c})$ gauge group}\\\\
Using the expressions $(3.12)$ and $(3.13)$, we find that:
\begin{equation} f_{UV}=\frac{15}{4}\left[\frac{N_{c}(N_{c}-1)}{2}+N_{f}N_{c}\right]\end{equation}
\begin{equation}f_{IR}=\frac{15}{4}\left[\frac{(2N_{f}-N_{c}+4)^{2}}{2}+\frac{N_{c}-4}{2}\right]\end{equation}
and the ACS inequality becomes:
\begin{equation}\frac{(2N_{f}-N_{c}+4)^{2}}{2}+\frac{N_{c}-4}{2}\leq\frac{N_{c}(N_{c}-1)}{2}+N_{f}N_{c}\end{equation}
\begin{equation} N_{f}\leq\frac{3}{2}(N_{c}-2)\end{equation}
Surprisingly again we find exactly the same condition as determined by the analysis of Seiberg for weak magnetic phase. Even here the inequality continue to holds in the $N_{f}>\frac{3}{2}(N_{c}-2)$ region as proved on section 2.
\\\\\textit{A.3. $Sp(2N_{c})$ gauge group}\\\\
Lastly we consider theory with $Sp(2N_{c})$ gauge group. Here we find with the aid of $(3.14)$ and $(3.15)$ that:
\begin{equation}f_{UV}=\frac{15}{4}\left[N_{c}(2N_{c}+1)+4N_{f}N_{c}\right]\end{equation}
\begin{equation}f_{IR}=\frac{15}{4}\left[2(2N_{f}-N_{c}-2)^{2}-N_{c}-2\right]\end{equation}
and the ACS inequality becomes:
\begin{equation}\frac{15}{4}\left[2(2N_{f}-N_{c}-2)^{2}-N_{c}-2\right]\leq\frac{15}{4}\left[N_{c}(2N_{c}+1)+4N_{f}N_{c}\right]\end{equation}
\begin{equation}N_{f}\leq\frac{3}{2}(N_{c}+1)\end{equation}
Which again corresponding to Seiberg analysis, while analysis on $N_{f}>\frac{3}{2}(N_{c}+1)$ appears on $7.2$ and show that it continue to hold.
\newpage
\section{References}
[1] T. Appelquist, A. G. Cohen, M. Schmaltz, A New Constraint on Strongly Coupled Field Theories,
Phys.Rev.D60:045003 (1999) [arXiv:hep-th/9901109v2].
\newline\newline[2] J.L. Cardy, Is There a c Theorem in Four-Dimensions?, Phys.Lett. B215, 749 (1988).
\newline\newline[3] A. B. Zamolodchikov Irreversibility of the flux of the renormalization group in a 2-D field theory JETP Lett. 43
\newline\newline[4] N. Seiberg, Electric - magnetic duality in supersymmetric non Abelian gauge theories,Nucl. Phys. B 435, 129 (1995)[arXiv:hep-th/9411149].
\newline\newline[5] K. A. Intriligator and N. Seiberg, Lectures on supersymmetric gauge theories and
electric-magnetic duality, Nucl. Phys. Proc. Suppl. 45BC, 1 (1996) [arXiv:hepth/9509066].
\newline\newline[6] V. A. Novikov, M. A. Shifman, A. I. Vainshtein and V. I. Zakharov, Exact Gell-MannLow Function Of Supersymmetric Yang-Mills Theories From Instanton Calculus,
Nucl. Phys. B 229, 381 (1983).
\newline\newline[7] D. Kutasov, A. Parnachev and D. A. Sahakyan, Central charges and U(1)R symmetries in N = 1 super Yang-Mills, JHEP 0311, 013 (2003) [arXiv:hep-th/0308071].
\newline\newline[8] K. Intriligator and B. Wecht, The exact superconformal R-symmetry maximizes a,
Nucl. Phys. B 667, 183 (2003) [arXiv:hep-th/0304128].
\newline\newline[9] K. Intriligator and B. Wecht, RG Fixed Points and Flows in SQCD with Adjoints,
Nucl.Phys.B677:223-272,2004 [arXiv:hep-th/0309201v1].
\newline\newline[10] M. E. Machacek and M. T. Vaughn, Two Loop Renormalization Group Equations In
A General Quantum Field Theory. 1. Wave Function Renormalization, Nucl. Phys.
B 222, 83 (1983).
\newline\newline[11] J. Grundberg, T. H. Hansson, and U. Lindstrom, Thermodynamics of
N=1 Supersymmetric QCD, hep-th/9510045
\newline\newline[12] J. I. Kapusta, Finite Temperature Field Theory, Cambridge University Press 1989.
\newline\newline[13] P. Arnold and C. Zhai The three-loop free energy for pure gauge QCD
[arXiv:hep-ph/9408276v1].
\newline\newline[14] Z. Komargodski, The Constraints of Conformal Symmetry on RG Flows, [arXiv:1112.4538v1]
\newline\newline[15] Z. Komargodski, A. Schwimmer On Renormalization Group Flows in Four Dimensions, [arXiv:1107.3987v2]
\newline\newline[16] K. A. Intriligator, R. G. Leigh and M. J. Strassler, New examples of duality in
chiral and nonchiral supersymmetric gauge theories, Nucl. Phys. B 456, 567 (1995)
[arXiv:hep-th/9506148].
\newline\newline [17] D. Anselmi, D. Z. Freedman, M. T. Grisaru and A. A. Johansen, Nonperturbative
formulas for central functions of supersymmetric gauge theories, Nucl. Phys. B 526,
543 (1998) [arXiv:hep-th/9708042].
\newline\newline[18] D. Anselmi, J. Erlich, D. Z. Freedman and A. A. Johansen, Positivity constraints
on anomalies in supersymmetric gauge theories, Phys. Rev. D 57, 7570 (1998)
[arXiv:hep-th/9711035].
\newline\newline[19] C. Csaki, P. Meade and J. Terning, A mixed phase of SUSY gauge theories from
a-maximization, JHEP 0404, 040 (2004) [arXiv:hep-th/0403062].
\newline[20] E. Barnes, K. Intriligator, B. Wecht and J. Wright, Evidence for the strongest version
of the 4d a-theorem, via a-maximization along RG flows, arXiv:hep-th/0408156.
\newline\newline[21] K. Intriligator and B. Wecht, Baryon charges in 4D superconformal field theories and
their AdS duals, Commun. Math. Phys. 245, 407 (2004) [arXiv:hep-th/0305046].
\newline\newline[22] K. Intriligator and B. Wecht, RG fixed points and flows in SQCD with adjoints,
Nucl. Phys. B 677, 223 (2004) [arXiv:hep-th/0309201].
\newline\newline[23] S. Martin, M. T. Vaughn Two-Loop Renormalization Group Equations for Soft Supersymmetry-Breaking Couplings
[arXiv:hep-ph/9311340v4]
\newline\newline[24] S. P. Martin A Supersymmetry Primer,
[arXiv:hep-ph/9709356v6]
\newline\newline[25] T. Banks, A. Zaks, On the phase structure of vector-like gauge theories with massless fermions Nucl. Phys. B196 (1982) 189.
\newline\newline[26] N. Craig, R. Essig, A. Hook, G. Torroba, New dynamics and dualities in supersymmetric chiral gauge theories, 
[arXiv:1106.5051v1].
\newline\newline[27] N. Craig, R. Essig, A. Hook, G. Torroba, Phases of N=1 supersymmetric chiral gauge theories, [arXiv:1110.5905v1].
\newline\newline[28] G. Mack, All unitary ray representations of the conformal group SU(2,2) with positive energy, Comm. Math. Phys. 55 (1977) 1.
\newline\newline[29] C. Zhai, B. Kastening, The Free Energy of Hot Gauge Theories with Fermions Through $g^{5}$, Nucl. Phys. B 628, 3 (2002) [arXiv:hep-th/0110028].
\newline\newline[30] M. Berkooz, The Dual of supersymmetric SU(2k) with an antisymmetric tensor and
composite dualities, Nucl. Phys. B 452, 513 (1995) [arXiv:hep-th/9505067].
\newline\newline[31] D. Kutasov, New results on the a-theorem in four dimensional supersymmetric field
theory, arXiv:hep-th/0312098.
\newline\newline[32] J. Grundberg, T.H. Hansson, U. Lindstrom, Thermodynamics of N=1 supersymmetric QCD, [arXiv:hep-th/9510045].
\newline\newline[33] D. Kutasov, A Comment on duality in N=1 supersymmetric nonAbelian gauge theories, Phys. Lett. B 351, 230 (1995) [arXiv:hep-th/9503086].
\newline\newline[34] D. Kutasov and A. Schwimmer, On duality in supersymmetric Yang-Mills theory,
Phys. Lett. B 354, 315 (1995) [arXiv:hep-th/9505004].
\newline\newline[35] D. Kutasov, A. Schwimmer and N. Seiberg, Chiral Rings, Singularity Theory and
Electric-Magnetic Duality, Nucl. Phys. B 459, 455 (1996) [arXiv:hep-th/9510222].
\newline\newline[36] K. A. Intriligator, New RG fixed points and duality in supersymmetric SP(N(c)) and
SO(N(c)) gauge theories, Nucl. Phys. B 448, 187 (1995) [arXiv:hep-th/9505051].
\newline\newline[37] R. G. Leigh and M. J. Strassler, Exactly marginal operators and duality in fourdimensional N=1 supersymmetric gauge theory, Nucl. Phys. B 447, 95 (1995)
[arXiv:hep-th/9503121].
\newline\newline[38] E. Witten, Barions in the $1/N$ expansion, Nuclear Physics B160 (1979) 57-115
\end{document}